\documentclass[hyper]{JHEP} 

\usepackage{epsfig}

\newcommand\fverb{\setbox\pippobox=\hbox\bgroup\verb}

\newcommand\fverbdo{\egroup\medskip\noindent%

            \fbox{\unhbox\pippobox}\ }

\newcommand\fverbit{\egroup\item[\fbox{\unhbox\pippobox}]}

\newbox\pippobox

\title{Hamiltonian Analysis of Non-Linear 
Sigma Model on Supercoset Target}

\author{J. Kluso\v{n}
 \footnote{On leave from Masaryk University, Brno}\\
Dipartimento di Fisica,\\
Universita' di Roma \& I. N. F. N Sezione di Roma 2, Tor Vergata \\
Via della Ricerca Scientifica, 1 00133  Roma   ITALY\\
E-mail:
\email{Josef.Kluson@roma2.infn.it}}
\preprint{ \\
\hepth{0608146}}
\abstract{This paper is devoted to the
study of the Hamiltonian formulation of 
non-linear sigma models on supercoset targets. We calculate
the Poisson brackets of left-invariant 
 currents. Then we introduce the Hamiltonian
of the system  
and determine the equations of motion for
left-invariant currents. We also determine
the charge corresponding to the invariance
of the action under global left multiplication.}


\def\tr{\mathrm{Tr}}

\def\str{\mathrm{Str}}

\def\pb  #1{\left\{#1\right\}}
\newcommand{\be}{\begin{equation}}
\newcommand{\ee}{\end{equation}}

\newcommand{\com}[1]{\left[#1\right]}

\newcommand{\mL}{\mathcal{L}}

\newcommand{\bg}{\mathbf{g}}
\newcommand{\bh}{\mathbf{h}}

\newcommand{\tJ}{\tilde{J}}

\newcommand{\mJ}{\mathcal{J}}
\begin{document}
\section{Introduction}
There has been recently  great
interest in non-linear sigma models on 
supercoset targets. These models
have many applications in several
branches of theoretical physics.
For example, they can be used in the
descriptions of the theory of 
quantum Hall efect
\cite{Guruswamy:1999hi,Read:2001pz}. 
However the main motivation for the
study of these models comes from 
string theory. 
In string theory the 
interest in the non-linear sigma
models on supercoset targets is
based on the remarkable observation that
these models give description
of the string theory in curved 
 Ramond-Ramond backgrounds
\footnote{For recent interestion
discussion of properties of
 supercosets background, see 
\cite{Grassi:2006cd}.}.
The Green-Schwarz action in $AdS_5
\times S_5$ formulated by Metsaev
and Tseytlin in \cite{Metsaev:1998it}
or its alternative form proposed and
further developed in 
\cite{Roiban:2000yy,Hatsuda:2002hz,
Hatsuda:2001xf} takes the form
of the sigma model on   
the coset superspace 
$PSU(2,2|4)/SO(1,4)\times SO(5)$.
 It was also soon discovered
that the sigma model on simpler
targets, the supergroups $PSL(n,n)$
has very rich and interesting structure
\cite{Bershadsky:1999hk,Berkovits:1999im,
Berkovits:1999zq}. In particular,
it was shown that sigma models on
cosets $G/H$ where $G$ is Ricci flat
supergroup and $H$ is its bosonic
subgroup are conformal to one loop -
and it is expected to be conformal
exactly - given a suitable Wess-Zumino term.
The important property of this construction
is that the isotropy subgroup $H$
is fix point set of  $Z_4$ grading
of $G$. 

It was further shown in \cite{Bena:2003wd}
that the $Z_4$ grading is a key element
for demonstration that the sigma model on
$PSU(2,2|4)/SO(1,4)\times SO(5)$ is classically
integrable
\footnote{The same conclusion was 
reached in \cite{Chen:2005uj} where
the flat currents  of the Green-Schwarz
superstrings in $AdS_5\times S^1$ and
$AdS_3\times S^3$ were constructed.}.
 This fact was also demonstrated
in case of the pure spinor formulation
of superstring on $AdS_5\times S_5$ 
\cite{Vallilo:2003nx,Berkovits:2004jw}. 
Moreover, since it was shown in
\cite{Berkovits:2004xu,Vallilo:2002mh}
 that the pure spinor string on
 $AdS_5\times S_5$ is consistent
 quantum theory one can hope that the
 integrability persists in the
 quantum regime as well. 
 
 It was shown recently in two
  remarkable papers 
 \cite{Kagan:2005wt,Young:2005jv}
that the similar constructions hold
 for sigma models on
spaces $G/H$ with more general
 $Z_{2n}$ grading.
For any $n$, the grading permits
an introduction of  certain preferred
Wess-Zumino term. An existence
of this  term implies  that 
the  equations of motion can be put 
into the Lax form ensuring the integrability
\cite{Young:2005jv}. It was then
shown in \cite{Kagan:2005wt}
 that these models are 
conformally invariant up to one loop.

Due  to  these facts we mean that it is
important to continue in the analysis
of these sigma models.
 In particular, the knowledge
of the classical Hamiltonian formalism
of non-linear sigma  models
on the supercoset targets
with $Z_{2n}$ grading
could be useful for further
study of these models.
For that reason the formulation of the 
the classical Hamiltonian analysis of
the sigma model on supercoset targets
is the main goal of this paper. 

Let us outline the structure of the paper.
In  section (\ref{second}) we introduce
the sigma model action on supercoset targets
following \cite{Kagan:2005wt} and we 
briefly review its basic properties. Then
in section (\ref{third}) we formulate
the Hamiltonian formalism of given theory.
We introduce  canonical variables
using the method proposed in  
\cite{Faddeev:1987ph}
\footnote{For recent application of this method
in string theory and quantum gravity,
see \cite{Das:2004hy,Das:2005hp,
Bianchi:2006im,Korotkin:1997fi,Miller:2006bu}.}.
In section (\ref{thirdP})
we calculate   the Poisson bracket
of the left-invariant currents. We check the validity of 
our calculations by comparing 
the Poisson bracket calculated here with the
Poisson bracket  derived in 
\cite{Bianchi:2006im} and we find
exact agreement.  
 In section (\ref{fourth}) we introduce
the Hamiltonian for   given system and also
determine the equations of motion of the
left-invariant 
currents. Finally, in section
(\ref{fifth}) we determine the charge corresponding
to the invariance of the action under global
left multiplication. We will
calculate the Poisson brackets between
these charges and spatial components
of the left-invariant currents. 

In summary, we mean that the properties of the
non-linear sigma model on supercoset targets
are very interesting  and certainly deserve
to be studied further. In particular, we
 hope that the results
derived  here could be helpful for an  analysis
of the quantum properties of these models. 
 It would be also certainly
very interesting to see whether the
supercosets with $Z_{2n}, n\neq 2$ grading
correspond to some
nontrivial background in string theory.  

\section{Formulation of non-linear sigma
model on supercoset targets}\label{second}
We begin this section with the brief review
of properties of supercosets. We do not
want to give the complete outline of this subject.
We rather focus on properties of the supercoset
that are necessary for definition of the
non-linear sigma model on it.

Let us consider  an associative Grassmann
algebra $\Lambda=\Lambda_0+\Lambda_1$,
where $\Lambda_0$ (resp. $\Lambda_1$)
consists of commuting (resp. anticommuting)
elements. Given a supermatrix $X=\left(\begin{array}{cc}
A & B \\
C & D \\ \end{array}\right)$ that belongs
to the supergroup $G$ we define it to
be even (odd) if $A,D\in \Lambda_0 \ 
(\Lambda_1)$ and $B,C\in \Lambda_1
\ (\Lambda_0)$. We introduce the
notation
 $|X|\equiv \mathrm{deg}(X)$
and we write $|X|=0$ if $X$ is
even matrix and $|X|=1$ if it
is odd matrix.  Then we can
define the supertrace as
\begin{equation}
\str(X)=\tr(A)-(-1)^{|X|}\tr D \ .
\end{equation}
This supertrace satisfies an
important property
\begin{equation}
\str (XY)=(-1)^{|X||Y|}
\str (YX) \ . 
\end{equation}

The relation between a supergroup
and superalgebra is similar to 
the bosonic case. The supergroup
$G$ associated to the superalgebra
$\bg$ is the exponential mapping
of the even subsuperalgebra 
of Grassmann envelope $\Lambda\otimes \bg$
\begin{equation}
g=\exp (x^AT_A) \ , 
\end{equation}
where $T_A$ generate the 
Lie superalgebra $\bg=\bg^{even}+
\bg^{odd}$ corresponding to 
$G$. We denote $|A|$ the
grade of $T_A$ in the Lie superalgebra.
Then $T_A$ satisfy the supercommutation
relationship 
\begin{equation}
[T_A,T_B]=T_AT_B-(-1)^{|A||B|}T_BT_A=f_{AB}^CT_C
\ ,
\end{equation}
where the structure constraints
$f_{AB}^C$ satisfy the graded Jacobi
identity
\begin{equation}
0=(-1)^{|A||C|}f_{AD}^E f_{BC}^D+
(-1)^{|B||A|}f^E_{BD}f^D_{CA}+
(-1)^{|C||B|}f^E_{CD}f^D_{AB} \ . 
\end{equation}
The $x^A$ commute (resp. anticommute) 
whenever $T_A$ are graded even (resp. odd).



As the next step let 
us now presume that $\bg$ is 
$Z_{2n}$ graded and $\bh$, the complexified
Lie algebra of the isotropy subgroup $H$
is the subspace of grade zero
\footnote{The $Z_{2n}$ grading is
defined by automorphism $\Omega:
\mathbf{g}\rightarrow \mathbf{g}$ such that
$[\Omega(X),\Omega(Y)]=\Omega([X,Y]) \ ,
\Omega^{2n}=1$ and $\Omega^k\neq 1$ for
all $k<2n$.}. That is
we suppose that $\bg$ may be written as
a direct sum
\begin{equation}
\bg=\bg_0+\bg_1+\dots +\bg_{2n-1}
\end{equation}
of vector subspaces where $\bg_0=\bh$ and that 
this decomposition respects the graded
Lie bracket
\begin{equation}\label{grlb}
\com{\bg_r,\bg_s}\subset  
\bg_{r+s \ mod \ 2n} \ . 
\end{equation}
We  presume that this grading is
compatible with the splitting to 
Grassmann odd and even variables, 
namely
\begin{equation}
\bg_{2s}\subset  \bg^{even} \ , \
\bg_{2s+1} \subset \bg^{odd} \  ,  \
\mathrm{for \ s}=0,1,\dots,n-1 \ .
\end{equation}
The basis $T_A$ can be chosen
to be a disjoint union of bases
 of $\bg_r$.  The basis element
 of $\bg_{r}$ will be denoted
 as $T_{i_{r}}$ where $i_r=
 1\dots,\mathrm{dim} \ \bg_{r}$.
We also use following conventions for the
naming of indices:
\begin{eqnarray}
i_A,j_B,\dots \ \mathrm{generators \ of} \ \bg \ , 
\nonumber \\
i_0\equiv x, j_0\equiv y,\dots \ \mathrm{generators \ of}
\ \bh \ , \nonumber \\
i_r,j_r,\dots \ \mathrm{generators \ of} \
\bg_{r} \ , \nonumber \\
 r=1,\dots,2n-1 \ . \nonumber \\
\end{eqnarray}
We also presume that the supertrace 
respect  the grading in a sense 
\footnote{This can be easily 
demonstrated in case when 
 the action of $\Omega$ on $X$
can be represented as an operation
of conjugation $\Omega(X)=MXM^{-1}$ for
some matrix $M$.}
 \begin{equation}\label{bilform}
\str XY=\str\Omega(X)\Omega(Y) \ . 
\end{equation}
Let us now presume that 
 $X\in \bg_r \ ,Y \in \bg_s$ 
so that 
\begin{equation}
\Omega(X)=e^{\frac{ir\pi}{n}}
 X \  ,
\Omega(Y)=e^{\frac{is\pi}{n}} X \  . 
\end{equation}
Then using  (\ref{bilform})
we immediately obtain 
\begin{equation}
\str XY=0 \ \mathrm{unless \ 
r+s=0 \ \mathrm{mod} \ 
2n \ .}
\end{equation}
Now we are ready to formulate the sigma
model on $G/H$, following 
\cite{Kagan:2005wt}. 
  We  express the sigma
model on $G/H$ in terms of a dynamical field
$g(x^\mu)\in G$ where $x^\mu, \  \mu=0,1$
 are worldsheet
coordinates. We write
\begin{equation}
J_\mu=g^{-1}\partial_\mu g \in \bg \ .
\end{equation}
Note that this current is invariant
under the global left action
\begin{equation}
g'=Ug \ , U\in G
 \ .
\end{equation}
We can decompose the current
into currents of  defined  grade
\begin{equation}
 J_\mu=J^{(0)}_\mu+
 \tJ_\mu \ , 
J^{(0)}_\mu \in \mathbf{h} \ , 
\tJ_\mu \in \bg/\bh
\ ,
\end{equation}
where
\begin{equation}
J^{(0)}_\mu=J_\mu^x T_x \ ,  \ 
\tJ_\mu=\sum_{r=1}^{2n-1}\sum_{i_r=1}^{\mathrm{dim} \ \bg_r}
J_\mu^{i_r}T_{i_r}\equiv
\sum_{r=1}^{2n-1}J^{(r)}_\mu \ .
\end{equation}
In what follows we use the Einstein
summation convention. Explicitly,
we define 
\begin{eqnarray}
\sum_{r=1}^{2n-1}\sum_{i_r=1}^{\mathrm{dim} \ \bg_r}
J_\mu^{i_r}T_{i_r}&\equiv&
\sum_r J_\mu^{i_r}T_{i_r} \ , \nonumber \\
\sum_{A=0}^{2n-1}\sum_{i_A=1}^{\mathrm{dim} \ \bg_A}
J_\mu^{i_A}T_{i_A}&\equiv&
\sum_A J_\mu^{i_A}T_{i_A} \  , 
i_0\equiv x \ .  \nonumber \\
\end{eqnarray}
Let us now study the transformation
properties of the current 
under the local right action $g'=gh \ , 
h(x)\in H$. Using the definition
of $J$ we easily obtain
\begin{eqnarray}
J'_\mu=h^{-1}J_\mu h+h^{-1}\partial_\mu h
 \ . 
\nonumber \\
\end{eqnarray}
Then using the fact that $\Omega(h)=h$
it is easy to determine
the rules how currents 
$J^{(0)}$ and $\tJ$ transform
under local right action  
\begin{eqnarray}
J'^{(0)}_\mu &=& h^{-1}\partial_\mu h+
h^{-1}J_\mu^{(0)} h \ , \nonumber \\
\tJ'_\mu  & =& h^{-1}\tJ_\mu h  \ . 
\nonumber \\
\end{eqnarray}
The next important object that
is needed for definition of the
non-linear sigma model on supercoset
target is 
 the metric 
\begin{equation}\label{metric}
K_{i_Aj_B}=\str (T_{i_A}T_{j_B}) \ . 
\end{equation}
This metric has   the  nonzero components 
\begin{equation}\label{graM}
K_{i_r j_{2n-r}}=(-1)^{|r|}
K_{j_{2n-r}i_r} \  , \ K_{xy}=K_{yx}   
\end{equation}
as follows from  (\ref{bilform}). 
Using  (\ref{graM})
we  define two form
field $B$ as 
\begin{equation}\label{Bfield}
B_{i_r j_{2n-r}}=q_r K_{i_r j_{2n-r}}
\ , q_{2n-r}=-q_r \ .
\end{equation}
Note also that (\ref{Bfield}) 
obeys the graded antisymmetry property
\begin{equation}
B_{i_r j_{2n-r}}=-(-1)^{|r|}
B_{j_{2n-r}i_r} \  
\end{equation}
that follows from (\ref{graM}) and
from the fact that $q_{2n-r}=-q_r$. 
Then we can write the action 
for non-linear sigma model on supercoset
target 
in the
form  
\begin{equation}\label{actbas}
S=-\int d^2x
\sum_r\left(\frac{1}{2}\eta^{\mu\nu}K_{i_r j_{2n-r}}
J^{i_r}_\mu J^{j_{2n-r}}_\nu
+\frac{1}{2}\epsilon^{\mu\nu} B_{i_rj_{2n-r}}
J^{i_r}_\mu J^{j_{2n-r}}_\nu
\right) \  .
\end{equation}
Equivalently, using   
(\ref{metric}) and (\ref{graM})
the action (\ref{actbas}) 
 can be written  as 
\begin{equation}\label{actbasstr}
S=-\int d^2x
\sum_r\str\left(\frac{1}{2}\eta^{\mu\nu}
J^{(r)}_\mu J^{(2n-r)}_\nu
+\frac{1}{2}\epsilon^{\mu\nu}
q_r J^{(r)}_\mu J^{(2n-r)}_\nu
\right) \  . 
\end{equation}
\section{Hamiltonian formalism}
\label{third}
Next step is to determine the canonical
variables and define corresponding
conjugate momenta. To do this we
follow  the approach introduced  in 
\cite{Faddeev:1987ph} 
. We start with the fact that
the current $J_\mu$ is flat
\begin{equation}
\partial_\mu J_\nu-\partial_\nu J_\mu+
\com{J_\mu,J_\nu}=0 \ .
\end{equation}
With the help of 
this identity we can
express $J_0$ as a function of $J_1$ if
we define  the operator $D$ as
\begin{equation}
\partial_0 J_1=\partial_1 J_0+
[J_1,J_0]\equiv D J_0 \ .
\end{equation}
Now we presume that the currents
$J_\mu$ obey appropriate boundary 
conditions so that we can introduce
the inverse operator $D^{-1}$ to
express $J_0$ as
\begin{equation}\label{relJ0J1}
J_0=D^{-1}(\partial_0 J_1)  \ . 
\end{equation}
If we insert (\ref{relJ0J1}) into the
action (\ref{actbasstr}) we obtain
the action that contain the dynamical variable
$J_1$ and its time  derivative
$\partial_0 J_1$. Then using  the
standard method we can define
the momenta conjugate to $J_1$.
 More precisely, let us extract the 
part of the action (\ref{actbasstr})
that contains the  time components of the
currents  
\begin{eqnarray}\label{S0}
 S_0=\frac{1}{2}\int d^2x\sum_r\str
 (J^{(r)}_0 J^{(2n-r)}_0-
 g_{r}J^{(r)}_0 J^{(2n-r)}_1)=
 \frac{1}{2}\int d^2x(\tJ_0 \tJ_0
 +\tJ_0 \hat{J}_1)=
\nonumber \\
=\frac{1}{2}\int d^2x\left[\tilde{D^{-1}(\partial_0 J_1)}
\tilde{D^{-1}(\partial_0 J_1)}
+\tilde{D^{-1}(\partial_0 J_1)} \hat{J}_1
\right] \ ,
\nonumber \\
\end{eqnarray}
where in the second step
we have used  (\ref{bilform})
and where we have also defined
\begin{equation}
\hat{J}_1= \sum_r q_{2n-r}J^{2n-r}_1 \ . 
\end{equation}
Finally the formula 
$\tilde{D^{-1}(\partial_0 J_1)}$ 
means that  we project to the 
  superalgebra of the coset $\bg/\bh$.
Then it is simple task to define
the momentum $\Pi_J$ as a variation of
the action (\ref{S0})
with respect to  $\partial_0 J_1$ 
and we get  
\begin{equation}\label{PiJ}
\Pi_J=\frac{\delta S}{\delta \partial_0 J_1}=
-\tilde{D^{-1}(D^{-1}(\partial_0J_1)+
\hat{J}_1)} \ , 
\end{equation}
where we have used the fact that
\begin{equation}
\int d^2x\str (D^{-1}(\delta X)(\dots))=
-\int d^2x \str 
(\delta X D^{-1}(\dots)) \ . 
\end{equation}
If we now act with $D$  on 
(\ref{PiJ}) from the left we obtain 
\begin{equation}\label{DPij}
D\Pi_J=-\tilde{D^{-1}
(\partial_0J_1)}-
\hat{J}_1
\end{equation}
that allows us to express
 $\tJ_0$ as function of canonical
 variables $\Pi_J$ and $J_1$
 \begin{equation}\label{j0pij}
\tJ_0=- D\Pi_J-
\hat{J}_1 \ . 
\end{equation}
To proceed  further 
 we expand $J_\mu$ and
 $\Pi_J$ as 
\begin{equation}\label{EXP}
\Pi_J=\sum_A \Pi^{i_A}T_{i_A} \ , 
\tJ_0=\sum_r J_0^{i_r}T_{i_r} \ ,
J_1=\sum_A J_1^{i_A}T_{i_A} \ , 
\hat{J}_1=\sum_r
q_rJ^{i_r}T_{i_r} \ .
\end{equation}
Then $D\Pi_J$ takes the form
\begin{eqnarray}\label{DPJex}
D\Pi_J=\partial_1\Pi_J+
\com{J_1,\Pi_J}=\nonumber \\
\sum_A(\partial_1 \Pi^{i_A}T_{i_A}+
 \sum_B J^{j_B}_1 \Pi^{k_{A-B}}f_{j_B k_{A-B}}^{i_A}T_{i_A})
\end{eqnarray}
using the fact that
the structure constants 
have   the form
$f_{i_A j_B}^{k_{A+B}}$ as
follows from (\ref{grlb}).
Then with the help of 
(\ref{DPJex}) and (\ref{EXP}) 
the equation (\ref{j0pij})
is equal to 
\begin{eqnarray}\label{Jir0}
-J^{i_r}_0 &=&
\partial_1 \Pi^{i_r}+\sum_A
J_1^{j_A}\Pi^{k_{r-A}}f_{j_Ak_{r-A}}^{i_r}+
q_r J^{i_r}_1 \ , 
\nonumber \\
\Phi^x &=& \partial_1\Pi^x+\sum_A
J_1^{i_A}\Pi^{j_{2n-A}}f_{i_A j_{2n-A}}^x
\approx 0 \ , 
\nonumber \\
\end{eqnarray}
where the absence of $J_0^x$ in
the action implies an existence
of the  primary constraint
$\Phi^x$. 

Let us now introduce the
 equal-time graded Poisson bracket that for
two classical observables $F,G$ depending 
on the phase super-space  variables
$ J_1^{i_A},\Pi_{j_A}$
 is defined as
\begin{eqnarray}
\pb{F,G} =(-1)^{|F||A|} \sum_A\left[\frac{\partial^L F}{\partial J_1^{i_A}}
\frac{\partial^L G}{\partial \Pi_{i_A}} -
(-1)^{|A|}\frac{\partial^L
F}{\partial \Pi_{i_A}} \frac{\partial^L G}{\partial J_1^{i_A}}\right] \ ,
\end{eqnarray}
where the superscript $L$ denotes left derivation.
 For the components 
$J_1=\sum_A J^{i_A}_1T_{i_A}$,
 $\Pi_J= \sum_A \Pi^{i_A}T_{i_A}=
\sum_A K^{i_Aj_{2n-A}}
\Pi_{j_{2n-A}}T_{i_A}$, the above PB's read
\begin{equation}\label{defPB}
\pb{J^{i_A}_1(x),\Pi_{j_B}(y)}=
(-1)^{|A|}\delta^A_B\delta^i_j
\delta(x-y) \  ,  
\end{equation}
where $\Pi_{i_A}$ is defined as
\begin{equation}
\Pi_{i_A}=
K_{i_A j_{2n-A}}\Pi^{j_{2n-A}} \ .
\end{equation}
We  again stress  that $A$ that
 labels the graded subspaces
$\bg^{(A)}$ is  in the one to one correspondence
with Grassmann property of given
elements, namely  $A$  odd
labels  Grassmann odd subalgebra and
$A$ even labels the Grassmann 
even subalgebra. 

Let us now define
\begin{equation}\label{defJit}
J^\mu_{i_t}\equiv K_{i_r j_{2n-r}}
J^{j_{2n-r}}_\mu \  . 
\end{equation} 
Then we can rewrite 
(\ref{Jir0}) in an alternative  form 
\begin{eqnarray}\label{DefJ0}
J^0_{i_r} &=&
-\partial_1 \Pi_{i_r}-\sum_A
 J^{k_A}_1 f_{i_r k_A}
^{j_{r+A}}\Pi_{j_{r+A}}+q_{r}J_{i_r}^1 \ , 
 \nonumber \\
\Phi_x &=& \partial_1\Pi_x+\sum_A
J^{k_A}_1f_{x k_A}^{j_A}
\Pi_{j_A} \  \nonumber \\
\end{eqnarray}
using the fact that $q_r=-q_{2n-r}$. 


\section{Calculation of the current algebra}
\label{thirdP}
In this section we will calculate the Poisson
brackets between currents $J_{\mu}^{i_A}$
using the canonical Poisson brackets
derived in the previous section. 
We start with the Poisson bracket
of the form $\pb{J^0_{i_r}(x),J_1^{j_s}(y)}$.
These Poisson brackets can be easily
calculated with the help of 
 (\ref{defPB}) and
(\ref{DefJ0}) and we   
obtain
\begin{eqnarray}\label{j01j}
\pb{J^0_{i_t}(x),J_1^{j_t}(y)} &=&
\delta_i^j\partial_x
\delta(x-y)+
 f_{i_t x}^{j_t}J^x_1(x)
\delta(x-y) \ , \nonumber \\
\pb{J^0_{i_t}(x),J_1^{j_r}(y)}&=&
f_{i_t l_{r-t}}^{j_r}J_1^{l_{r-t}}(x)
\delta(x-y) \ , 
\mathrm{for \ r\neq t} \ , 
\nonumber \\
\pb{J^0_{i_t}(x),J_1^x(y)} &=&
f_{i_tj_{2n-t}}^xJ_1^{j_{2n-t}}(x)
\delta(x-y) \ . 
\end{eqnarray}
Using (\ref{defJit})
we obtain the alternative
form of  Poisson bracket
(\ref{j01j}) 
\begin{eqnarray}\label{J0iupJ1}
\pb{J^{i_{2n-t}}_0(x),
J^{j_t}(y)}&=& K^{i_{2n-t}j_t}
\partial_x\delta(x-y)+J^x_1(x)
f_{x k_{2n-t}}^{i_{2n-t}}
K^{k_{2n-t}j_t} \  ,
\nonumber \\
\pb{J^{i_t}_0(x),J^{j_r}_1(y)}&=&
J^{l_{r+t}}_1(x)f_{l_{r+t}k_{2n-r}}
^{i_t}K^{k_{2n-r}j_r} \ ,
\nonumber \\ 
\pb{J^{i_t}_0(x),J_1^x(y)} &=&
J_1^{j_{t}}(x)f_{j_{t}y}^{i_t}K^{yx}
\delta(x-y) \  ,  
\nonumber \\
\end{eqnarray}
where we have also used the fact
that $2n+t\sim t$. 

In the same way we can determine
the Poisson bracket
between $\Phi_x$ and $J^{i_r}_1,
J^y_1$
\begin{eqnarray}
\pb{\Phi_z(x),J^{v}_1(y)}&=&
-\delta^v_z\partial_x\delta(x-y)
-f_{zw}^vJ^w_1(x)\delta(x-y) \  ,
\nonumber \\
\pb{\Phi_x(x),J^{i_r}_1(y)}&=&
-f_{x j_r}^{i_r}J_1^{j_r}(x)\delta(x-y)   
 \ . \nonumber \\  
\end{eqnarray}
These Poisson brackets 
demonstrate how currents $J_1^{i_r}$
transform  under the 
gauge transformations generated by
$\Phi_x$.  
It is also clear that $J_0^{i_r}$ 
has to transform in the same
way so that 
\begin{equation}
\pb{\Phi_x(x),J^{i_r}_0(y)}=
-f_{x j_r}^{i_r}J_0^{j_r}(x)
\delta(x-y) 
\end{equation}
or equivalently
\begin{equation}
\pb{\Phi_x(x),J_{i_r}^0(y)}=
J^0_{j_r}(x)f_{x i_r}^{j_r}
\delta(x-y)  \ . 
\end{equation}

Now we will calculate the
Poisson brackets between zero components
of the currents $J_0^{i_A}$.  
Let us start with the Poisson bracket 
$\pb{J^0_{i_t}(x),
J^0_{j_{2n-t}}(y)}$. 
After straightforward, but
slightly involved  calculation 
we  obtain
\begin{equation}\label{i00}
\pb{J^0_{i_t}(x),J^0_{j_{2n-t}}(y)}=
(-1)^{|t|}\Phi_x(x)
f_{i_tj_{2n-t}}^x\delta(x-y) \  ,
\end{equation}
where we have used
  (\ref{defPB}) and
(\ref{DefJ0}) and also the
fact that the structure
functions of the supercoset
obey  the relation 
\begin{equation}
f_{i_Aj_B}^{k_{A+B}}
 K_{k_{A+B}l_C}=
-(-1)^{|B||C|}f_{i_A l_C}^{k_{A+C}}
K_{k_{A+C}j_B} \   
\end{equation}
and  the graded
Jacobi identity 
\begin{equation}
0=(-1)^{|A||C|}f_{i_Aj_{B+C}}^{m_{A+B+C}}
 f_{k_Bl_C}^{j_{B+C}}+
(-1)^{|B||A|}f^{m_{A+B+C}}_{k_Bj_{A+C}}
f^{j_{A+C}}_{l_Ci_A}+
(-1)^{|C||B|}f^{m_{A+B+C}}_{l_Cj_{A+B}}
f^{j_{A+B}}_{i_Ak_B} \ , 
\end{equation}
where we have also used  (\ref{grlb}).
Finally note  that (\ref{i00})
can be written in the form 
\begin{equation}\label{J0J0up}
\pb{J^{i_{2n-t}}_0(x),
J^{j_t}_0(y)}=-\Phi^x(x) 
f_{x k_{2n-t}}^{i_{2n-t}}
 K^{k_{2n-t}i_t} \ , 
\end{equation}
where $\Phi^x=\Phi_y K^{yx}$. 

As the next step we will calculate
the Poisson bracket  
\begin{equation}\label{j0itjs}
\pb{J^0_{i_t}(x),J^0_{j_s}(y)} \ ,
t\neq s \ .  
\end{equation}
This Poisson bracket
 can be calculated exactly
in the same way as in the previous
cases however now the
result strongly depends on the
value of  $q_t$ 
that appears in (\ref{Bfield})  
and that according to 
 \cite{Kagan:2005wt,Young:2005jv}
is equal to 
\begin{equation}
q_s=1-\frac{s}{n} \ . 
\end{equation}
More precisely it turns out
that for $t+s<2n$ the Poisson
bracket (\ref{j0itjs})
takes the form 
\begin{equation}\label{j0itjs1}
\pb{J^0_{i_t}(x),J^0_{j_s}(y)}=
-(-1)^{|s||t|}(J_{l_{t+s}}^1+
J_{l_{t+s}}^0)(x)f^{l_{t+s}}_{i_tj_s}
\delta(x-y) \ 
\end{equation}
while for   $t+s>2n$ it is equal to 
\begin{equation}\label{j0itjs2}
\pb{J^0_{i_t}(x),J^0_{j_s}(y)}=
(-1)^{|s||t|}(J_{l_{t+s}}^1-
J_{l_{t+s}}^0)(x)f^{l_{t+s}}_{i_tj_s}
\delta(x-y) \  .
\end{equation}
It will be also useful to express
(\ref{j0itjs1}) and (\ref{j0itjs2})
in the alternative form as
\begin{eqnarray}\label{J0J0up1}
\pb{J_0^{i_t}(x),J_0^{j_s}(y)}&=&
(J_0^{l_{t+s}}+J_1^{l_{t+s}})(x)
f_{l_{t+s}k_{2n-s}}^{i_t}K^{k_{2n-s}j_s}\delta(x-y) \ , \
\mathrm{for \ t+s>2n} \ , \
\nonumber \\  
\pb{J_0^{i_t}(x),J_0^{j_s}(y)}&=&
(J^{l_{t+s}}_0-
J^{l_{t+s}}_1)(x)f_{l_{t+s}k_{2n-s}}^{i_t}
K^{k_{2n-s}j_s} 
\delta(x-y) \ , \  \mathrm{for \ t+s<2n}
\ . \nonumber \\
\end{eqnarray} 
We would like to stress that
 the case 
 $n=2$ was previously studied
in the context 
of the pure spinor superstring  on
$AdS_5\times S_5$. The algebra
of left-invariant currents was
also  determined in 
\cite{Bianchi:2006im}. 
Then it is easy to see that 
the Poisson brackets
(\ref{J0J0up1}) 
coincide with the  
Poisson brackets derived there
if we omit the contributions
of the  ghost fields. This agreement  
serves as further justification
of our result. 

\section{Hamiltonian and equations of motion}
\label{fourth}
In this section we introduce of the
Hamiltonian for the non-linear sigma
model on supercoset target. 
Using the action given
in (\ref{actbasstr}) we obtain the
matter part of the Hamiltonian in the
form 
\begin{equation}\label{H}
H_{matt}=\int dx \str (\partial_0J
\Pi-\mL)=
\frac{1}{2}\int dx \str(J_0J_0+
J_1J_1)
\end{equation}
or alternatively 
\begin{eqnarray}\label{H1}
H_{matt}&=&\frac{1}{2}\int dx \sum_r\left(J^{i_r}_0K_{i_rj_{2n-r}}
J^{j_{2n-r}}_0+J^{i_r}_1 K_{i_r j_{2n-r}}
J^{j_{2n-r}}_1\right)
\nonumber \\
&=&\frac{1}{2}
\int dx \sum_r\left((-1)^{|r|}
J_{i_r}^0K^{i_rj_{2n-r}}
J_{j_{2n-r}}^0+
(-1)^{|r|}J_{i_r}^1 K^{i_r j_{2n-r}}
J_{j_{2n-r}}^1\right) \ . 
\nonumber \\
\end{eqnarray}
Following the general theory
of constraint systems we have
to introduce to the Hamiltonian
the contribution that corresponds to the
fact that the dynamics of the system 
is restricted on the constraint surface $\Phi_x=0$
\begin{equation}
H_{con}=\int dx \Gamma^x \Phi_x(x) \ . 
\end{equation}
It can be shown that the
time evolution of the primary
constraints $\Phi_x$ 
 does not induce
any secondary constraints so that
the whole Hamiltonian takes the form
\begin{equation}\label{Htot}
H=H_{matt}+H_{con} \ . 
\end{equation}
With the help of the Hamiltonian
(\ref{Htot}) we can determine 
the equations of motion for 
 $J_{i_r}^0,J^{i_s}_1$ using the
fact that  the time evolution of
any function $X(J,\Pi)$ that is  defined 
on the phase space spanned by
$J_1^{i_A},\Pi_{j_B}$ is
governed by the equation
\begin{equation}
\partial_0 X=\pb{X,H} \ .  
\end{equation}
With the help of 
 the Poisson bracket that were derived
in the previous section and the form 
of the Hamiltonian given above we
immediately obtain the equation
of motion for  
 $J^{i_s}_1$ in the form 
\begin{eqnarray}\label{J0ir}
\partial_0 J^{i_r}_1=
\partial_x J_0^{i_{r}}
+J_1^x J^{j_r}_0 f_{x j_r}^{i_r}
-\sum_t J_0^{j_t}J_1^{l_{r-t}}f_{j_t
l_{r-t}}^{i_r}+\Gamma^x J_1^{j_r}f_{x j_r}^{i_r} \ .
\nonumber \\
\end{eqnarray}
The form of the equation above suggests
that it is natural to choose the Lagrange
multiplier $\Gamma^x$ to be equal to
(In other words we fix the gauge)
\begin{equation}\label{Gammax}
\Gamma^x=-J^x_0 \  
\end{equation}
 and introduce
the covariant derivative
\begin{eqnarray}
\nabla_\mu X^{i_r}_\nu &\equiv&
\partial_\mu X^{i_r}_\nu+
J^x_\mu X^{j_r}_\nu f_{x j_r}^{i_r} \ ,
\nonumber \\ 
\nabla_\mu X^\nu_{i_r}
&\equiv&
\partial_\mu X^\nu_{i_r}-J^x X_{j_r}^\nu f^{j_r}_{x
i_r} \ , \nonumber \\
\nabla_\mu X^\nu_{i_r}&=&K_{i_r j_{2n-r}}\nabla_\mu 
X^{j_{2n-r}}_\nu \  . \nonumber \\ 
\end{eqnarray}
Then we  can rewrite the equation (\ref{J0ir})
 into the form
\begin{equation}\label{dj1}
-\nabla_0 J_1^{i_r}+\nabla_1 J^{i_r}_0
-\sum_t
J^{j_t}_0J^{l_{r-t}}_1
 f_{j_t l_{r-t}}^{i_t}=0 \ . 
\end{equation}
In the same way we can determine
the equation of motion for 
$J^0_{i_r}$  
\begin{eqnarray}\label{Jir0a}
&-& \nabla_0 J_{i_r}^0+
\nabla_1 J_{i_r}^1-\sum_t
(-1)^{|r||t|+|t|}J_{l_{r+t}}^0 J_0^{j_t}
f_{i_r j_t}^{l_{r+t}}-\nonumber \\
&-& \sum_{r+t<2n}(-1)^{|t|+|r||t|}J_{l_{r+t}}^1 J_0^{j_t}
f_{i_r j_t}^{l_{r+t}}
+\sum_{r+t>2n}(-1)^{|t|+|r||t|}J_{l_{r+t}}^1 J_0^{j_t}
f_{i_r j_t}^{l_{r+t}}+
\nonumber \\
&+&\sum_{r+t\neq 2n}J_1^{l_{t-r}}f_{i_r l_{t-r}}^{j_t}
K_{j_t k_{2n-t}}J^{k_{2n-t}}_1 
=0 \ , 
\nonumber \\
\end{eqnarray}
where we have also used 
(\ref{Gammax}).  Alternatively,
using the currents $J_\mu^{i_r}$
we can rewrite (\ref{Jir0a}) into 
the form 
\begin{eqnarray}\label{djoe}
&-&\nabla_0 J^{i_{r}}_0+
\nabla_1 J^{i_{r}}_1 
+\sum_t J^{k_{r-t}}_0J^{l_t}_0
f_{k_{r-t}l_t}^{i_{r}}+
\sum_{r+t\neq 2n}J_1^{k_{r-t}}
J^{l_t}_1 
f_{k_{r-t}l_t}^{i_r}
-\nonumber \\
&-&\sum_{t-r<0}
J_1^{j_t}J_0^{k_{r-t}}
f_{j_t k_{r-t}}^{i_{r}}+
\sum_{t-r>0}
J_1^{j_t}J_0^{k_{r-t}}
f_{j_t k_{r-t}}^{i_r}
=0 \ . 
\nonumber \\
\end{eqnarray}
The form of the  equation (\ref{djoe})
simplifies for $r=1$ and for $r=2n-1$. 
In case $r=1$  it is natural to
 combine  (\ref{dj1}) with
 (\ref{djoe}) and we obtain 
\begin{eqnarray}\label{eqjr1}
(\eta^{\mu\nu}-\epsilon^{\mu\nu})
\nabla_\mu J^{i_1}_\nu+
(\eta^{\mu\nu}-\epsilon^{\mu\nu})
\sum_{t=2}^{2n-2}
J_\mu^{k_{2n+1-t}}
J^{l_t}_\nu 
f_{k_{2n+1-t}l_t}^{i_1}=0 \ . 
\nonumber \\
\end{eqnarray}
On the other hand for $r=2n-1$ we
 subtract 
(\ref{dj1}) from   (\ref{djoe})
and we get
\begin{eqnarray}
(\eta^{\mu\nu}+\epsilon^{\mu\nu})
\nabla_\mu J^{i_{2n-1}}_\nu+
(\eta^{\mu\nu}+\epsilon^{\mu\nu})
\sum_{t=1}^{2n-2}
J_\mu^{j_{t}}J_\nu^{k_{2n-1-t}}
f_{j_t k_{2n-1-t}}^{i_{2n-1}}=0 \ . 
\end{eqnarray}
For arbitrary  $r$ we can perform
the same manipulation. We
firstly add (\ref{dj1}) to (\ref{djoe})
and we obtain 
\begin{eqnarray}
(\eta^{\mu\nu}-\epsilon^{\mu\nu})
\nabla_\mu J^{i_r}_\nu+
\eta^{\mu\nu}\sum_{r+t\neq 2n}
J_\mu^{k_{r-t}}
J^{l_t}_\nu 
f_{k_{r-t}l_t}^{i_r}
\nonumber \\
-\sum_{r+t\neq 2n}
J_0^{k_t}J_1^{l_{r-t}}f_{k_t l_{r-t}}^{i_r}
-\sum_{t-r<0}
J_1^{k_t}J_0^{l_{r-t}}
f_{k_t l_{r-t}}^{i_{r}}+
\sum_{t-r>0}
J_1^{k_t}J_0^{l_{r-t}}
f_{k_t l_{r-t}}^{i_r}=0 \ . 
\nonumber \\
\end{eqnarray}
On the other hand if we
subtract 
 (\ref{djoe}) from (\ref{dj1})
we get 
\begin{eqnarray}
(\eta^{\mu\nu}+\epsilon^{\mu\nu})
\nabla_\mu J^{i_r}_\nu+
\eta^{\mu\nu}\sum_{r+t\neq 2n}
J_\mu^{k_{r-t}}
J^{l_t}_\nu 
f_{k_{r-t}l_t}^{i_r}
\nonumber \\
+\sum_{r+t\neq 2n}
J_0^{k_t}J_1^{l_{r-t}}f_{k_t l_{r-t}}^{i_r}
-\sum_{t-r<0}
J_1^{k_t}J_0^{l_{r-t}}
f_{k_t l_{r-t}}^{i_{r}}+
\sum_{t-r>0}
J_1^{k_t}J_0^{l_{r-t}}
f_{k_t l_{r-t}}^{i_r}=0 \ . 
\nonumber \\
\end{eqnarray}
It is easy to see that for $n=2$ these
equations of motion 
coincide with the equations
of motion derived in 
\cite{Bianchi:2006im} when we
again ommit the contributions
of ghost fields. 

We conclude this section
with the derivation of the
equation of motion for $J_1^x$.
With the help of the Poisson
brackets derived in 
section (\ref{thirdP}) we immediately
obtain 
\begin{eqnarray}
\partial_0 J_1^x(x)=
-\sum_t (-1)^{|t|}J_1^{j_{2n-t}} f_{i_t j_{2n-t} }^x
K^{i_t k_{2n-t}} J_{k_{2n-t}}^0
+J_1^w f_{yw}^x \Gamma^y-\partial_x
 \Gamma^x \ . 
\nonumber \\
\end{eqnarray}
Using (\ref{Gammax}) we can
rewrite this equation in the form 
\begin{equation}
\partial_0 J_1^x-\partial_1 J_0^x+\sum_t J_0^{i_t}
J_1^{j_{2n-t}}f_{i_t j_{2n-t}}^x+
J_0^y J_1^z f_{yz}^x=0
\end{equation}
that is  Maurer-Cartan
identity for $J^x_\mu$.  In other words the dynamics
of $J_\mu^x$ is trivial. 
\section{Global symmetry of the non-linear sigma
model}
\label{fifth}
By definition the
left-invariant 
currents  $J^{i_A}=(g^{-1}dg)^{i_A}$
(and consequently the action) 
are invariant 
under the global symmetry $g'=hg$ where
$h$ is a constant element from $G$.
 Our goal is  to determine
the corresponding conserved charge
using standard Noether procedure.
To do this let us presume that
$h\approx 1+\epsilon $, where
$\epsilon(x)=\epsilon^{i_A}T_{i_A}$
 depends on the worldvolume coordinates.
Then the variation of the current
is equal to
\begin{equation}
\delta J_\mu= g^{-1}\partial_\mu \epsilon g
\end{equation}
that implies the variations
of the components of the
currents $J^{i_A}_\mu$
in the form 
\begin{eqnarray}\label{deltaJA}
\delta J^{i_A}_\mu &=&
K^{i_Aj_{2n-A}}\str(g T_{j_{2n-A}} g^{-1}T_{k_C})
\partial_\mu\epsilon^{k_C} \nonumber \\
&=&
K^{i_Aj_{2n-A}}C_{j_{2n-A}k_C}\partial_\mu \epsilon^{k_C}
\ . \nonumber \\ 
\end{eqnarray}
It is important to stress that $C_{i_A j_B}$ is
Grassmann odd for $|A+B|=1$ and is Grassmann
even for $|A+B|=0$ as follows from the properties
of the supertrace and from the fact that the
generator $T_{i_A}$ is odd matrix for $|A|=1$. 

Now with the help of
  (\ref{deltaJA})
we obtain  that the  variation of the action
is equal to 
\begin{eqnarray}
\delta S=
-\int d^2x \sum_t(
J^{j_{2n-t}}_\mu(\eta^{\mu\nu}-
q_t \epsilon^{\mu\nu})
C_{j_{2n-t}i_A})\partial_\nu \epsilon^{i_A}
\equiv \int d^2x \partial_\nu \mathcal{J}^\nu_{i_A}
\epsilon^{i_A} \ . 
\nonumber \\
\end{eqnarray}
Since for fields that are  on-shell any variation of
the action has to vanish the
expression above  implies 
\begin{equation}\label{muja}
\partial_\mu \mathcal{J}^\mu_{i_A}=0
\ , 
\end{equation}
where
\begin{equation}
\mathcal{J}^{\mu}_{i_A}=
\sum_t(\eta^{\mu\nu}-
q_t \epsilon^{\mu\nu})J^{j_{2n-t}}_\nu 
C_{j_{2n-t}i_A} \ . 
\end{equation}
Using (\ref{muja}) we can 
 define the conserved  charge
\begin{equation}
q_{i_A}=\int dx \mJ^0_{i_A}=
-\int dx \sum_t
(J^{j_{2n-t}}_0+
q_t J^{j_{2n-t}}_1) 
C_{j_{2n-t}i_A} \ . 
\end{equation}
It is instructive to calculate 
the Poisson bracket between $q_{i_A}$
and $J_1^{j_B}$. With the help of 
the Poisson
brackets that were determined in
section (\ref{thirdP}) we obtain 
\begin{eqnarray}\label{deltaJq}
\pb{J^{i_t}_1(x),q_{j_B}}&=&
-K^{i_t k_{2n-t}}\partial_x
C_{k_{2n-t} j_B}-
J^{l_{t-r}}_1 (x)
f_{l_{t-r}m_{r}}^{i_t}
K^{m_{r}k_{2n-r}}
C_{k_{2n-r}j_B} \ , 
\nonumber \\ 
\pb{J^x_1(x),q_{j_B}}&=&
-J^{l_{2n-r}}_1 (x)
f_{l_{2n-r}m_{r}}^{x}
K^{m_{r}k_{2n-r}}
C_{k_{2n-r}j_B} \ . \nonumber \\
\end{eqnarray} 
We can  simplify this result using
the definition of $C_{i_Aj_B}$ and
the properties of left-invariant
currents since 
\begin{eqnarray}\label{partialxC}
\partial_x C_{i_Bj_C}
&=&\str (\partial_x gT_{i_B} g^{-1}T_{j_C})-
\str (T_{i_B} q^{-1}\partial_x gg^{-1}T_{j_C} g)=
\nonumber \\
&=&\str (g^{-1}\partial_x g T_{i_B} g^{-1}T_{j_C} g)-
(-1)^{|B||C|}\str (g^{-1}\partial_x g g^{-1}T_{j_C}gT_{i_B})=
\nonumber \\
&=&\sum_A J^{k_A}\str ([T_{k_A}T_{i_B}
-(-1)^{|A||B|}T_{i_B}T_{k_A}]g^{-1}T_{j_C} g)=
\nonumber \\
&=& \sum_A J^{k_A}f_{k_A i_B}^{l_D}
\str (T_{l_D} g^{-1} T_{j_C} g)=
\sum_A J^{k_A} f_{k_Ai_B}^{l_D} 
C_{l_Dj_C} \ .  \nonumber \\
\end{eqnarray}
With the help of this result
the first equation in (\ref{deltaJq})
takes the form 
\begin{eqnarray}
\pb{J^{i_t}_1(x),q_{j_B}}&=&
\sum_A (-1)^{|t||t+A|}
J^{l_{t-A}}_1 f_{l_{t-A}m_{A}}^{i_t}
K^{m_{A}k_{2n-A}}C_{k_{2n-A} j_B}(x)
-\nonumber \\
&-& \sum_r J^{l_{t-r}}_1 
f_{l_{t-r}m_{r}}^{i_t}
K^{m_{r}k_{2n-r}}
C_{k_{2n-r}j_B}(x) \ . 
\nonumber \\ 
\end{eqnarray}
For $|t|=0$ the equation  above 
simplifies as 
\begin{eqnarray}\label{t0}
\pb{J^{i_t}_1(x),q_{j_B}}=
J^{l_{t}}_1 f_{l_{t}x}^{i_t}
K^{xy}C_{y j_B}=
\delta_{gauge}J^{i_t}_1(x)(C_{y j_B}(x))
\ . 
\nonumber \\
\end{eqnarray}
In other words 
 the Poisson 
bracket between $J_1^{i_t},|t|=0$ and
$q_{j_B}$ is equal to the
 gauge transformation of 
current $J^{i_t}_1$ with the gauge
parameter equal to $C_{y j_B}(x)$. However for
$|t|=1$  we do not obtain such a 
clear interpretation since
\begin{eqnarray}\label{ji2s}
\pb{J^{i_{2s+1}}_1(x),q_{j_B}}&=&
-2\sum_{t=1}^{n-1}
J^{l_{2s+1-2t}}_1 f_{l_{2s+1-2t}m_{2t}}^{i_{2s+1}}
K^{m_{2t}k_{2n-2t}}C_{k_{2n-2t} j_B}(x)+
\nonumber \\
&+&J^{l_{2s+1}}_1 
f_{l_{2s+1}x}^{i_{2s+1}}
K^{xy}
C_{yj_B}(x) \  . 
\nonumber \\ 
\end{eqnarray} 
We again see that   the expression on the second
line in (\ref{ji2s})
can be written as $\delta_{gauge}
J^{i_{2s+1}}(C_{yj_B})$.
On the other hand it is not clear to 
us how to  interpret
 the expression 
on the first line.  

Finally, using (\ref{partialxC}) 
the second equation in (\ref{deltaJq})
can be written as 
\begin{equation}
\pb{J^x_1(x),q_{j_B}}=
\sum_r J^{l_{2n-r}}_1 (x)
f_{l_{2n-r}x}^{k_{2n-r}}
K^{xy}
C_{yj_B}=
\sum_r\delta_{gauge} J^{k_{2n-r}}_1
 (C_{y j_B}(x)) 
 \end{equation} 
that can be again interpreted as a
sum of the gauge transformations of 
currents $J^{i_t}_1$. 
At present it is not completely
clear to us how to interpret these
results. Since
the left-invariant currents are invariant
under the transformations $g'=hg$ 
one could expect that the Poisson brackets
between left-invariant  currents
and the charges corresponding to the global
symmetry of the action 
are either zero or equal to the
gauge transformations that define coset.
We have seen that this
interpretation holds for  Grassmann even
components of the 
left-invariant currents. On the other hand 
we have derived
different results 
for  the
Grassmann odd components of 
the  left-invariant 
currents  
and as we mentioned above
it is not clear to us how to interpret
this result. 
This issue certainly
deserve more precise study and we hope
to return to it in future.  

\section*{Acknowledgements}
We would like to thank M.~ Bianchi and  A.~Das
 for  useful discussions.   This work was supported in
part by INFN, by the MIUR-COFIN contract 2003-023852, by the EU
contracts MRTN-CT-2004-503369 and MRTN-CT-2004-512194, by the
INTAS contract 03-516346 and by the NATO grant PST.CLG.978785
and in part  by the Czech Ministry of
Education under Contract No. MSM 0021622409.



\begin{thebibliography}{20}



\bibitem{Guruswamy:1999hi}
  S.~Guruswamy, A.~LeClair and A.~W.~W.~Ludwig,
   \emph{
``Gl(N$|$N) Super-Current 
Algebras For Disordered Dirac Fermions In Two
Dimensions,''}
  Nucl.\ Phys.\ B {\bf 583} (2000) 475
  [arXiv:cond-mat/9909143].

\bibitem{Read:2001pz}
  N.~Read and H.~Saleur,
 \emph{``Exact spectra of conformal supersymmetric nonlinear sigma models in two
  dimensions,''}
  Nucl.\ Phys.\ B {\bf 613} (2001) 409
  [arXiv:hep-th/0106124].

\bibitem{Grassi:2006cd}
  P.~A.~Grassi and M.~Marescotti,
\emph{``Flux vacua and supermanifolds,''}
  arXiv:hep-th/0607243.








\bibitem{Metsaev:1998it}
  R.~R.~Metsaev and A.~A.~Tseytlin,
 \emph{``Type IIB superstring 
action in AdS(5) x S(5) background,''}
  Nucl.\ Phys.\ B {\bf 533} (1998) 109
  [arXiv:hep-th/9805028].



\bibitem{Roiban:2000yy}
  R.~Roiban and W.~Siegel,
 \emph{``Superstrings on 
AdS(5) x S(5) supertwistor space,''}
  JHEP {\bf 0011} (2000) 024
  [arXiv:hep-th/0010104].

\bibitem{Hatsuda:2002hz}
  M.~Hatsuda and M.~Sakaguchi,
  \emph{``Wess-Zumino term for AdS superstring,''}
  Phys.\ Rev.\ D {\bf 66} (2002) 045020
  [arXiv:hep-th/0205092].

\bibitem{Hatsuda:2001xf}
  M.~Hatsuda and K.~Kamimura,
 \emph{``Classical AdS superstring mechanics,''}
  Nucl.\ Phys.\ B {\bf 611} (2001) 77
  [arXiv:hep-th/0106202].


  
  
\bibitem{Bershadsky:1999hk}
  M.~Bershadsky, S.~Zhukov and A.~Vaintrob,
\emph{``PSL(n$|$n) sigma model as a conformal field theory,''}
  Nucl.\ Phys.\ B {\bf 559} (1999) 205
  [arXiv:hep-th/9902180].

\bibitem{Berkovits:1999im}
  N.~Berkovits, C.~Vafa and E.~Witten,
\emph{``Conformal field theory 
of AdS background with Ramond-Ramond flux,''}
  JHEP {\bf 9903} (1999) 018
  [arXiv:hep-th/9902098].
  
\bibitem{Berkovits:1999zq}
  N.~Berkovits, M.~Bershadsky, T.~Hauer, S.~Zhukov and B.~Zwiebach,
\emph{``Superstring theory on AdS(2) x S(2) 
as a coset supermanifold,''}
  Nucl.\ Phys.\ B {\bf 567} (2000) 61
  [arXiv:hep-th/9907200].

\bibitem{Bena:2003wd}
  I.~Bena, J.~Polchinski and R.~Roiban,
 \emph{``Hidden symmetries 
of the AdS(5) x S**5 superstring,''}
  Phys.\ Rev.\ D {\bf 69} (2004) 046002
  [arXiv:hep-th/0305116].

\bibitem{Chen:2005uj}
  B.~Chen, Y.~L.~He, P.~Zhang and X.~C.~Song,
\emph{``Flat currents of the 
Green-Schwarz superstrings in AdS(5) x S**1 and
AdS(3) x S**3 backgrounds,''}
  Phys.\ Rev.\ D {\bf 71}, 086007 (2005)
  [arXiv:hep-th/0503089].





\bibitem{Vallilo:2003nx}
  B.~C.~Vallilo,
\emph{``Flat currents in 
the classical AdS(5) x S**5 pure spinor superstring,''}
  JHEP {\bf 0403} (2004) 037
  [arXiv:hep-th/0307018].

\bibitem{Berkovits:2004jw}
  N.~Berkovits,
  \emph{``BRST cohomology 
and nonlocal conserved charges,''}
  JHEP {\bf 0502} (2005) 060
  [arXiv:hep-th/0409159].




\bibitem{Berkovits:2004xu}
  N.~Berkovits,
 \emph{``Quantum consistency of 
the superstring in AdS(5) x S**5 background,''}
  JHEP {\bf 0503} (2005) 041
  [arXiv:hep-th/0411170].

\bibitem{Vallilo:2002mh}
  B.~C.~Vallilo,
\emph{``One loop conformal 
invariance of the superstring in an AdS(5) x S(5)
background,''}
  JHEP {\bf 0212} (2002) 042
  [arXiv:hep-th/0210064].
  
   
\bibitem{Kagan:2005wt}
  D.~Kagan and C.~A.~S.~Young,
 \emph{``Conformal 
Sigma-Models On Supercoset Targets,''}
  Nucl.\ Phys.\ B {\bf 745} (2006) 109
  [arXiv:hep-th/0512250].
  
\bibitem{Young:2005jv}
  C.~A.~S.~Young,
\emph{``Non-Local Charges, 
Z(M) Gradings And Coset Space Actions,''}
  Phys.\ Lett.\ B {\bf 632} (2006) 559
  [arXiv:hep-th/0503008].


\bibitem{Faddeev:1987ph}
  L.~D.~Faddeev and L.~A.~Takhtajan,
 \emph{``HAMILTONIAN 
METHODS IN THE THEORY OF SOLITONS,''}


  
   
\bibitem{Das:2004hy}
  A.~Das, J.~Maharana, A.~Melikyan and M.~Sato,
  \emph{``The algebra of 
transition matrices for the AdS(5) x S**5 superstring,''}
  JHEP {\bf 0412}, 055 (2004)
  [arXiv:hep-th/0411200].
  
\bibitem{Das:2005hp}
  A.~Das, A.~Melikyan and M.~Sato,
\emph{``The algebra of flat currents for the string on AdS(5) x S**5 in the
  light-cone gauge,''}
  JHEP {\bf 0511}, 015 (2005)
  [arXiv:hep-th/0508183].
  
\bibitem{Bianchi:2006im}
  M.~Bianchi and J.~Kluson,
 \emph{``Current Algebra 
Of The Pure Spinor Superstring In Ads(5) X S(5),''}
  arXiv:hep-th/0606188.





\bibitem{Korotkin:1997fi}
  D.~Korotkin and H.~Samtleben,
 \emph{``Yangian symmetry in 
integrable quantum gravity,''}
  Nucl.\ Phys.\ B {\bf 527} (1998) 657
  [arXiv:hep-th/9710210].

\bibitem{Miller:2006bu}
  B.~H.~Miller,
 \emph{``Conserved charges in 
the principal chiral model on a supergroup,''}
  JHEP {\bf 0608} (2006) 010
  [arXiv:hep-th/0602006].





\end{thebibliography}
\end{document}